\documentclass[english,aps,prl,twocolumn,graphics,graphicx
]{revtex4-1}
\usepackage[utf8]{inputenc}

\usepackage{amsmath,mathtools}
\usepackage{amssymb}
\usepackage{graphicx}
\usepackage{verbatim}
\usepackage{braket}
\usepackage{soul}
\usepackage{bm}
\usepackage{gensymb}
\usepackage{diagbox}
\usepackage{siunitx}
\usepackage[T1]{fontenc}
\usepackage[english]{babel}
\usepackage{mathtools}
\usepackage{newtxtext}  
\usepackage{amsmath}
\usepackage{amsfonts}
\usepackage{hyperref}

\renewcommand\vec{\boldsymbol}
\begin{document}
\title{ 
 Two-Body Solution and Instabilities along Středa Lines in Moir\'e Flat Bands
}


\author{Guopeng Xu and Chunli Huang}
\affiliation{Department of Physics and Astronomy, University of Kentucky, Lexington, Kentucky 40506-0055, USA}
\date{\today} 

\begin{abstract}
Moir\'e  minibands in twisted homobilayer semiconductors can, under suitable approximations, be modeled as a pair of Landau levels with opposite Chern numbers. This provides a minimal model  for searching novel topological states in a time-reversal–symmetric Hamiltonian. In this work, we investigate the effects of an external magnetic field in this model.
We study the many-body ground state in the density–magnetic-field ($n$-$B$) plane along the $dn/dB=\pm1/\Phi_0$ Středa line with Hartree-Fock approximation. Away from charge neutrality, we find the Chern-insulating (incompressible) state is very robust while
towards charge neutrality, we find a transition from incompressible phase to compressible phase as the interaction strength $\kappa$ decreases. Using time-dependent mean-field theory, we further analyze spin-flip excitations and find that the incompressible state along St\v{r}eda line toward charge neutrality becomes unstable even at large $\kappa$ when magnetic field is sufficiently large.
Finally, we solve the two-body problem in a given Landau level exactly where the two particles experience unequal magnetic fields using a new basis called center-of-charge basis.
This basis allows any isotropic interaction to be parameterized by a single quantum number, the relative angular momentum, thereby extending the  Haldane pseudopotentials to the unequal-magnetic-fields case. As the difference of the two magnetic fields varies, these pseudopotentials show a sequence of level crossings, leading to non-monotonic structure of pseudopotentials that is absent in ordinary Landau level systems.
Our formulation provides a useful starting point for studying weak-field physics in moir\'e flat bands, where magnetic Bloch-state basis becomes computationally impossible due to the large basis sizes.
\end{abstract}
\maketitle

\section{Introduction}

Since the discovery of correlated states in moir\'e materials with flat minibands\cite{Cao2018,Cao2018SC,Guo_2025,Xia2025,xu2025signaturesunconventionalsuperconductivitynear,https://doi.org/10.7910/dvn/y5mtri,Lu2024,Tang2020,Wang2020,Shimazaki2020,Li2021,PhysRevB.103.125146,Regan2020,Xu2020,Huang2021,Jin2021,Tang2022,doi:10.1126/science.aaw3780,doi:10.1126/science.aay5533,Zeng2023,Cai2023,Park2023,PhysRevX.13.031037,Kang2024, kang2025timereversalsymmetrybreakingfractional}, some of which \cite{doi:10.1126/science.aaw3780,doi:10.1126/science.aay5533,Zeng2023,Cai2023,Park2023, PhysRevX.13.031037} spontaneously break time-reversal symmetry and exhibit (fractional) quantum anomalous Hall effects. There has been growing interest in identifying new classes of topological states that do not have direct analogs in conventional quantum Hall systems.
A convenient starting point is a model based on Landau levels. Landau levels form perfectly flat Chern bands and therefore provide an ideal platform for interaction-driven correlated states. A time-reversal-symmetric Hamiltonian can then be constructed by introducing an additional set of Landau levels with opposite chirality and including all symmetry-allowed interactions between them. This toy model and its variant have been studied previously by several groups \cite{zou2025valley, PhysRevLett.124.166601,PhysRevResearch.7.023083,kwan2021exciton,kwan2022excitonic,kwan2025textured}, including our own work \cite{xu2025localizedexcitonslandaulevelmixing}. In certain limits, such models can approximate the moir\'e minibands of twisted transition-metal dichalcogenides in the adiabatic limit \cite{PhysRevLett.132.096602}. A natural question is how such systems evolve when time-reversal symmetry is explicitly broken by an external magnetic field $B$. Can the magnetic field seed new topological correlated states? This question is also experimentally relevant, since magnetic fields are among the most commonly used probes in transport experiments. In this work, we investigate this problem and analyze the resulting correlated states and collective excitations.

Recent experiments in twisted MoTe$_2$ systems \cite{Cai2023,Zeng2023,Park2023,PhysRevX.13.031037} have reported a prominent incompressible gap (vanishing longitudinal resistance $R_{xx}$) in the density–magnetic-field phase diagram along the line $dn/dB=\pm 1/\Phi_0$  pointing away from charge neutrality at filling fraction $\nu=1$ per moir\'e unit cell. We refer to lines with slope $dn/dB=\pm 1/\Phi_0$ in the $n$–$B$ plane as St\v{r}eda lines, even though the incompressible state may not always be groundstate along these lines. In contrast, no comparable feature is observed along the line  toward neutrality. This asymmetry indicates that the thermodynamically favored state at strong magnetic field carries a Chern number $C=-\mathrm{sgn}(B)$ in the devices studied in Refs.~\cite{Cai2023,Zeng2023,Park2023,PhysRevX.13.031037}.

This experimental observation raises an important question: what is the microscopic mechanism that lead to this preference? While magnetization energy certainly provides one important energy scale, other contributions, including kinetic, exchange and correlation energy, also play a significant role. Ref.~\cite{PhysRevB.110.L201107} carried out an extensive self-consistent mean-field study of the Hofstadter spectrum in twisted MoTe$_2$, building on their earlier theoretical work \cite{wang2024theory,wang2022narrow}, see also \cite{kolavr2024hofstadter}. Their approach incorporates realistic model parameters and captures many experimental features. However, because the Hofstadter basis size grows rapidly as the magnetic field decreases, their calculations cannot be smoothly extended to the $B\to0$ limit. While our toy model does not capture all microscopic details, such as the moir\'e periodic potential, it has the advantage that it can be studied continuously down to zero magnetic field. By introducing a new basis, which we refer to as the center-of-charge  basis, we can extend both mean-field and time-dependent mean-field calculations continuously to $B=0$ and obtain analytical expressions for both the ground-state energy and the excitation spectrum.

There are both universal and model-dependent aspects to the energetics along the St\v{r}eda line. For the St\v{r}eda line pointing away from charge neutrality, the (kinetic) cyclotron energy disfavors the incompressible Chern-insulator state, while the exchange energy favors it. The situation is opposite for  the St\v{r}eda line pointing toward neutrality: there both the cyclotron energy and the exchange energy favor the incompressible Chern insulator. These trends represent the universal aspects of the problem. However, these universal considerations alone cannot account for the experimental observations in MoTe$_2$. An important model-dependent contribution is the spin and valley Zeeman energy. For parameters relevant to MoTe$_2$ \cite{PhysRevLett.122.086402}, we find that along the St\v{r}eda line away from neutrality the Zeeman energy favors the incompressible Chern-insulator state, whereas along the St\v{r}eda line toward neutrality it disfavors it. Taken together, these effects provide a natural explanation for the experimental observations \cite{Cai2023,Zeng2023,Park2023,PhysRevX.13.031037} and are consistent with previous theoretical results \cite{PhysRevB.110.L201107}. In addition, using time-dependent Hartree–Fock theory we compute the spin-flip excitation spectrum along the St\v{r}eda line and find that, at sufficiently large magnetic fields, the incompressible Chern-insulator state pointing toward charge neutrality becomes unstable.

The remainder of this manuscript is organized as follows. In Sec.~\ref{sec:Streda_line_groundstate}, we compute the energies of the incompressible Chern-insulator state and a competing compressible state within the Hartree–Fock approximation. In Sec.~\ref{sec:spin-flip}, we use the time-dependent Hartree–Fock method to compute the spin-flip exciton spectrum. In Sec.~\ref{sec:two_body_opposite}, we introduce the center-of-charge  basis and apply it to solve the two-body problem. Finally, in Sec.~\ref{sec:conclusion}, we discuss the implications of our results for recent experiments and outline possible directions for future work.
\begin{figure}
    \centering
    \includegraphics[width=1\linewidth]{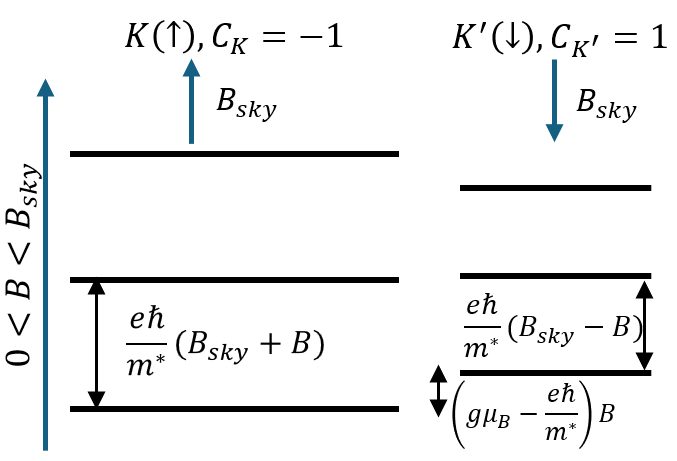}
    \caption{Single particle energy-level $\epsilon_{n\sigma}^0$ in Eq.~\ref{eq:single_particle_LL} and \ref{eq:single_particle_LL-2}. The applied magnetic field is aligned with Skyrmion magnetic field in  valley $K$ so the Landau-level degeneracy in valley $K$ is greater than in valley $K'$. The energy offset between the two valleys is determined by the competition of Zeeman splitting and the cyclotron energy of the applied field. 
    For $g>e\hbar/m^*c\approx 3$, the lowest energy level in valley $K$ is lower than that in valley $K'$, as shown above.
    }
\label{fig:schematic_LLs}
\end{figure}

\begin{figure*}
    \centering
    \includegraphics[width=1\linewidth]{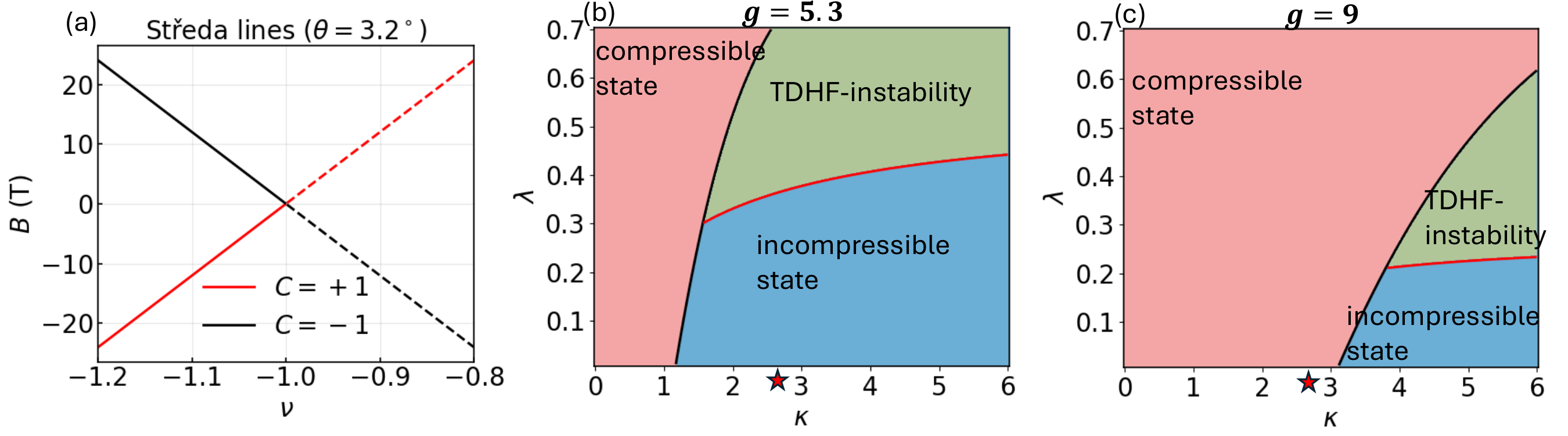}
    \caption{(a) Schematic illustration of the two St\v{r}eda lines emerging from filling fraction $\nu=-1$ in the $(\nu,B)$ plane. The solid line denotes the branch where an incompressible gap (vanishing longitudinal resistance $R_{xx}$) is observed experimentally, while the dashed line denotes the branch where no incompressible feature is observed \cite{Cai2023,Zeng2023,Park2023,PhysRevX.13.031037}.
(b–c) $\kappa-\lambda$ phase diagram along the St\v{r}eda line toward charge neutrality (red dashed line in panel (a)) for g-factor $g=5.3$ and $g=9$. The horizontal axis denotes the interaction strength $\kappa\approx92/(\theta\epsilon)$, and the vertical axis denotes the ratio of external magnetic field to the skyrmion field, $\lambda= B/B_{\mathrm{sky}}$.
The red region corresponds to the parameter regime where the compressible state $|\Psi_{CO}^T\rangle$ is the ground state. The blue region corresponds to the regime where the incompressible state $|\Psi_{IC}^T\rangle$ is the ground state. The green region indicates that the lowest spin-reversal excitation energy of $|\Psi_{IC}^T\rangle$ becomes negative.
The star on the $x$-axis indicates our estimated $\kappa\sim2.6$ for twisted MoTe$_2$ using dielectric constant $\epsilon=10$ and twisted angle $\theta=3.5^\circ$. 
}\label{fig:main_result}
\end{figure*}

\section{Hartree-Fock approximation along the two St\v{r}eda Lines}

\label{sec:Streda_line_groundstate}

The many body Hamiltonian we consider is the following,
\begin{align}\label{eq:many_body_H}
    H &= \sum_{nm} \epsilon_{n\uparrow}^0 c^\dagger_{nm\uparrow}c_{nm\uparrow} +\sum_{n\bar m}
    \epsilon_{n\downarrow}^0 c^\dagger_{n\bar m\downarrow}c_{n\bar m\downarrow} 
     \\
    &+ \frac{1}{2}\frac{e^2}{\epsilon l_0}
\sum_{12'34'}\sum_{\sigma\sigma'}
\bra{1\sigma,2'\sigma'}V_c\ket{3\sigma,4'\sigma'}
    c_{1\sigma}^{\dagger} c_{2'\sigma'}^{\dagger} c_{4'\sigma'} c_{3\sigma} \nonumber 
\end{align}
where the single-particle energy levels are given by the following,
\begin{align}\label{eq:single_particle_LL}
\epsilon^0_{n\uparrow}=\hbar\omega_0(1+\lambda)\left(n+\tfrac{1}{2}\right)-\frac{1}{2}g\mu_B  B\\
\epsilon^0_{n\downarrow}=\hbar\omega_0(1-\lambda)\left(n+\tfrac{1}{2}\right)+\frac{1}{2}g\mu_B B.\label{eq:single_particle_LL-2}
\end{align}
Here $\omega_0=eB_{sky}/m^*$ is the cyclotron frequency of the emergent Skyrmion field and is roughly $\hbar\omega_0\approx 2.1(\theta[\text{Degree}])^2 \text{ }$meV\cite{PhysRevLett.132.096602}.We use an overbar on the quantum numbers of the spin-$\downarrow$ states to indicate that the effective magnetic field they experience is opposite in direction to that of the spin-$\uparrow$ states, see Ref~\cite{xu2025localizedexcitonslandaulevelmixing}. We introduce a  parameter $ 0<\lambda<1$ to characterize the difference between the applied field $ B$ to the emergent Skyrmion field $B_{\text{sky}}$,
\begin{align} 
\lambda = \frac{ B}{B_{\text{sky}}} \approx 0.085 \frac{B[T]}{\theta[\text{Degree}]^2}
\end{align}
The Coulomb energy scale is given by $e^2/\epsilon l_0 \approx 193 \theta[\text{Degree}]/\epsilon  \text{ }meV$ where $l_0$ is the magnetic length from Skyrmion field. The $V$'s are (dimensionless) Coulomb interaction matrix elements and they depend on $\lambda$ through the single-particle wavefunctions. Here number "1" is a shorthand for $n_1m_1$ etc, see Appendix~\ref{sec:Coulomb matrix elements opposite spins} for more discussion on these matrix elements. We define the dimensionless parameter $\kappa$ as the ratio of Coulomb energy scale to the cyclotron energy:
\begin{align}\label{eq:kappa_def}
    \kappa = \frac{e^2}{\epsilon l_0 \hbar \omega_0} \approx \frac{92}{\theta[\text{Degree}] \epsilon}
\end{align}
For typical twisted angles and screening environment (hBN substrate), the $\kappa$ is large and can range from $2-7$.

The key physics controlling the system at finite magnetic field arises from the single-particle energies $\epsilon^0_{n\uparrow}$ and $\epsilon^0_{n\downarrow}$, which are illustrated schematically in Fig.~\ref{fig:schematic_LLs}. Two main effects occur. First, electrons with opposite spins experience opposite magnetic fields of different strengths. Their orbits are therefore Landau quantized by different effective fields: $\pm B_{\text{sky}}+B$. Consequently, the cyclotron gap is larger for spin-up electrons, which experience the larger field $B_{\text{sky}}+ B$, than for spin-down electrons, which experience the smaller field $-B_{\text{sky}}+B$. The Landau-level degeneracy (density of states) is also different for the two spin species:
\begin{align}\label{eq:imbalance degen}
    N_\uparrow &= \frac{(B_{\text{sky}} +  B)A}{\Phi_0} = (1+\lambda) N_\phi, \\
    N_\downarrow &= \frac{(B_{\text{sky}} -  B)A}{\Phi_0} = (1-\lambda) N_\phi,
\end{align}
where $N_\phi = \frac{B_{\text{sky}}A}{\Phi_0}$ is the number of states in the energy level without external magnetic field.

Second, there is an additional spin-dependent single-particle energy splitting arising from the Zeeman effect.  
In the presence of spin–orbit coupling \cite{PhysRevLett.122.086402}, the spin and valley degrees of freedom are locked for the low-energy physics considered here. The valence band in the $K$ valley (with orbital character $d_{x^2-y^2}+i d_{xy}$) corresponds to spin $\uparrow$, while the $K'$ valley (with orbital character $d_{x^2-y^2}-i d_{xy}$) corresponds to spin $\downarrow$. The valley $g$-factor has the same sign as the spin $g$-factor. Consequently, a magnetic field aligned with spin $\uparrow$ lowers the energy by $\tfrac{1}{2} g\mu_B B$, where the effective $g$ factor is given by 
\begin{equation}
    g=g_s + g_{\mathrm{valley}}
\end{equation} 
where $g_s \approx 2$ is the conventional spin Zeeman coupling and $g_{\mathrm{valley}}$ is the valley contribution.  
Ref.~\cite{PhysRevB.110.L201107} estimates the effective $g_{\mathrm{valley}}\sim3.2$, while other works suggest even larger values \cite{Zhang__2023}. The ratio of this spin splitting energy to the cyclotron energy is,
\begin{align}\label{eq:Zeeman_energy_scale}
    \frac{E_Z}{\hbar\omega_0} \approx 0.163g\lambda
\end{align}
For twisted angle  $\theta$ from $3^\circ$ to $4^\circ$, the Skymion field is about $30-40$Tesla, the typical $\lambda$ can be up to $0.2$. Consequently, the Zeeman energy is not negligible in this regime. Although we refer to this as Zeeman splitting, it actually includes contributions from both spin and valley degrees of freedom. We emphasize that the emergent Skyrmion magnetic field does not contribute to the Zeeman splitting; only the external magnetic field enters the Zeeman energy.

Let us note that the relation between valley and the Chern number of the moir\'e minibands (or equivalently the direction of the emergent skyrmion magnetic field) depends on the specific parameters of the moir\'e Hamiltonian \cite{PhysRevLett.132.036501,PhysRevLett.122.086402}. In Ref.~\cite{PhysRevLett.122.086402}, the first moir\'e miniband in the $K$ valley carries Chern number $C=-1$, while  the opposite valley has $C=+1$. Using the parameter set of Ref.~\cite{PhysRevLett.132.036501}, the assignment is reversed: the $K$-valley has $C=+1$, while the $K'$ valley has $C=-1$. In this work, we follow the convention of Ref.~\cite{PhysRevLett.122.086402}, which implies the correspondence
\begin{align}
C_K=-1 \leftrightarrow K \leftrightarrow \uparrow, \;\;\;
C_{K'}=+1 \leftrightarrow K' \leftrightarrow \downarrow .
\end{align}
For this reason, we will often use the labels $C$, valley, and spin interchangeably.
However, as pointed out in Refs.~\cite{PhysRevLett.122.086402,Zhang2024}, the sign of the Chern number of the moir\'e minibands is not in one-to-one correspondence with the sign of the skyrmion magnetic field. In particular, at sufficiently small twist angles the simple picture based on a single real-space skyrmion texture breaks down \cite{Zhang2024}. A proper description in this regime requires a more detailed microscopic treatment of the moir\'e potential and lattice relaxation effects, which lies beyond the scope of the present manuscript.

\subsection{$\nu=1$ St\v{r}eda Line Away From Charge Neutrality}
Without loss of generality, we take $B>0$ and focus on the St\v{r}eda line away from the charge neutrality, defined by $dn/dB=-1/\Phi_0$, shown as the black solid line in Fig.~\ref{fig:main_result}(a). 
The single-particle spectrum is shown in Fig.~\ref{fig:schematic_LLs}.
Along this line the degeneracy of the spin-$\uparrow$ Landau levels increases with magnetic field in precisely the way required to accommodate the change in electron density. Consequently, the electron density can always completely fill the lowest spin-$\uparrow$ Landau level.
The most natural state to consider is therefore the fully occupied spin-$\uparrow$ Landau level, which we refer to as the \textit{incompressible Chern insulator}. If this state is indeed the ground state along the St\v{r}eda line, it will lead to a resistance minimum and a quantized Hall response along the trajectory $dn/dB=-1/\Phi_0$ in the $n$–$B$ plane.

A possible competing state arises if electrons first occupy the spin-$\downarrow$ Landau level, which has a smaller cyclotron energy, see Fig.~\ref{fig:schematic_LLs}. Because the total electron density exceeds the number of available states in the spin-$\downarrow$ Landau level, the remaining electrons must ``spill over'' into the spin-$\uparrow$ Landau level. This leads to a partially filled spin-$\uparrow$ level which we refer to as a \textit{compressible state}.  To determine which state is energetically favored, we evaluate their energies within the Hartree-Fock approximation.

 \begin{table}[t]
\centering
\renewcommand{\arraystretch}{1.2}
\begin{tabular}{lcc}
\hline\hline
 & \multicolumn{2}{c}{St\v{r}eda line} \\
\cline{2-3}
Energy contribution 
& Away from CNP 
& Toward CNP \\
\hline
Kinetic (cyclotron) energy 
& $\times$ 
& $\checkmark$ \\

Zeeman energy 
& $\checkmark$
& $\times$ \\

Exchange energy
& $\checkmark$ 
& $\checkmark$ \\
\hline\hline
\end{tabular}
\caption{Summary of which energy contributions favor the incompressible Chern-insulator state along the two St\v{r}eda lines. $\checkmark$ ($\times$) indicates that the corresponding term favors (disfavors) the incompressible phase. The $g$ factor entering the Zeeman energy includes both valley and spin contributions, $g=g_{\text{valley}}+g_s$. When the skyrmion magnetic field reverses direction, the role of the Zeeman energy is interchanged between the two Středa lines. }
\label{tab:energy_competition}
\end{table}

For the incompressible Chern-insulator state $|\Psi_{IC}^{A}\rangle = \prod_{m}^{N_\uparrow} c_{0m\uparrow}^\dagger |0\rangle$, the Hartree–Fock energy per particle, measured in units of $\hbar\omega_0$, is $\varepsilon_{IC}^{A} = \frac{1}{N_\uparrow}\langle \Psi_{IC}^{A}|H|\Psi_{IC}^{A}\rangle$. A straightforward evaluation leads to
\begin{align}
     \varepsilon_{IC}^{A} 
     & = \frac{1+\lambda}{2} - \frac{\kappa}{2}\sqrt{\frac{\pi}{2}}\sqrt{1+\lambda} - 0.163g\lambda.
 \end{align}
This incompressible Chern insulator preserves $M_{xz}T$ symmetry since both mirror $M_{xz}$ and $T$  flip the sign of $B$ and is called the Chern paraelectric in Ref.~\cite{wang2024theory}. The first term in $\varepsilon_{IC}^{A}$ represents the cyclotron energy of the zeroth Landau level and increases with $\lambda$. The second term is the exchange energy. It modifies the usual exchange energy of a completely filled lowest Landau level, $-\frac{\kappa}{2}\sqrt{\frac{\pi}{2}}$, by a factor of $\sqrt{1+\lambda}$. This enhancement arises because the Landau-level wavefunctions become more localized as $\lambda$ increases, see Sec.~\ref{sec:two_body_opposite}. The third term corresponds to the Zeeman energy, expressed in units of $\hbar\omega_0$.

 For the compressible state, denoted as $| \Psi_{CO}^A\rangle$, we assume it is spatially uniform, and the density matrix takes the following form:
\begin{align}\label{eq:compressible_state}
     \langle \Psi_{CO}^A|c^\dagger_{m\uparrow}c_{m'\uparrow}|\Psi_{CO}^A\rangle &= \delta_{mm'}\frac{N_\uparrow-N_\downarrow}{N_{\uparrow} }= \delta_{mm'} \frac{2\lambda}{1+\lambda}, \nonumber \\
     \langle \Psi_{CO}^A|c^\dagger_{m\downarrow}c_{m'\downarrow}|\Psi_{CO}^A\rangle &= \delta_{mm'},
 \end{align}
 The energy per particle $\varepsilon_{CO}^A = \frac{1}{N_\uparrow}\langle \Psi_{CO}^A|H|\Psi_{CO}^A\rangle$ is evaluated in Sec.~\ref{sec:energy_per_prticles} and the result is,
 \begin{align}
    \varepsilon_{CO}^A 
     &= \frac{1+3\lambda^2}{2(1+\lambda)}
        - \frac{\kappa}{2}\sqrt{\frac{\pi}{2}}\left[
             \frac{4\lambda^2}{(1+\lambda)^{3/2}} + \frac{(1-\lambda)^{3/2}}{1+\lambda}
          \right] \notag\\
          & - 0.163\frac{3\lambda^2-\lambda}{1+\lambda} g.
 \end{align}
Comparing the two energies, we found $\varepsilon_{IC}^{A} <\varepsilon_{CO}^{A} $ for  $g>0$ and $\kappa \gtrsim0.6$. This is the parameter regime relevant to our system. Physically, the exchange energy favors the incompressible state because it maximizes the number of electrons with the same spin projection. The Zeeman energy of the incompressible state is also lower than that of the compressible state because all electrons in the incompressible state align with the direction favored by the external magnetic field. The only energy that favors the compressible state is the cyclotron-energy because the cyclotron gap of the spin-$\downarrow$ Landau level is smaller. However, as discussed earlier, realistic estimates of the $g$ factor satisfy $g > e\hbar/(m^*\mu_B)$, implying that the Zeeman splitting dominates over the cyclotron-energy scale. Consequently, the incompressible Chern insulator is the ground state along this St\v{r}eda line. A summary of which energy contributions favor each state is provided in Table~\ref{tab:energy_competition}.

As we will show below, this conclusion does not hold for the St\v{r}eda line pointing towards the charge neutrality.


\subsection{$\nu=1$ St\v{r}eda Line Toward Charge Neutrality}

 Along the St\v{r}eda line toward charge neutrality, carriers are removed in such a way that the spin-$\downarrow$ Landau level, whose degeneracy decreases with magnetic field (recall $B>0$), can remain completely filled. A fully occupied spin-$\downarrow$ Landau level therefore represents the incompressible state along this trajectory and the corresponding Slater determinant is
\begin{align}\label{eq:incompressible_state}
    |\Psi_{IC}^{T}\rangle &= \prod_m c^\dagger_{0m \downarrow}|0\rangle
\end{align}
The Hartree--Fock energy per particle (in units of $\hbar\omega_0$), $ \varepsilon_{\mathrm{IC}}^T
= \frac{1}{N_\downarrow} \langle \Psi_{IC}^T|H|\Psi_{IC}^T\rangle$, can be evaluated and yields,
\begin{align}
    \varepsilon_{\mathrm{IC}}^T=
\frac{1-\lambda}{2}
-
\frac{\kappa}{2}\sqrt{\frac{\pi}{2}}\sqrt{1-\lambda}
+
0.163\,g\lambda
\end{align}
A natural competing configuration is the compressible state $|\Psi_{CO}^T\rangle$ in which the $N_\downarrow$ electrons partially occupy the spin-$\uparrow$ Landau level. This level contains $N_\uparrow$ orbitals with $N_\uparrow > N_\downarrow$. Assuming a spatially uniform state, the density matrix satisfies the form
\begin{align}
    \langle \Psi_{CO}^T|c^\dagger_{\uparrow m}c_{\uparrow m'}|\Psi_{CO}^T\rangle =\frac{N_{\uparrow}}{N_{\downarrow}}\delta_{mm'}=\frac{1-\lambda}{1+\lambda}\delta_{mm'}
\end{align}
The Hartree-Fock energy per particle of this compressible state is $\varepsilon_{CO}^T
=\frac{1}{N_\downarrow}\langle \Psi_{CO}^T|H|\Psi_{CO}^T\rangle $,
\begin{align}
\varepsilon_{CO}^T
&=\frac{1+\lambda}{2}
-
\frac{\kappa}{2}\sqrt{\frac{\pi}{2}}
\,\frac{1-\lambda}{\sqrt{1+\lambda}}
-
0.163g\lambda
\end{align}
Along this St\v{r}eda branch (T), the two energy levels $\varepsilon_{CO}^T$ and $\varepsilon_{IC}^T$ can cross depending on the parameters $\kappa$, $g$, and $\lambda$. In Fig.~\ref{fig:main_result}(b–c), we plot the resulting phase diagram along this St\v{r}eda branch. The $x$-axis represents the dimensionless interaction strength $\kappa$, while the $y$-axis denotes the dimensionless magnetic field $\lambda$. Results are shown for two representative values of the $g$ factor, $g=5.3$ and $g=9$. 

The red region indicates the parameter regime in which the compressible state has lower energy than the incompressible state, $\varepsilon_{CO}^T < \varepsilon_{IC}^T$. The blue and green regions correspond to the regime where the incompressible state has lower energy, $\varepsilon_{IC}^T < \varepsilon_{CO}^T$. A first-order phase transition line separates these two phases.
The green region further indicates that the incompressible state becomes unstable within time-dependent Hartree–Fock theory, as discussed in Sec.~\ref{sec:spin-flip}.

For small $\kappa$, the exchange energy is weak compared to the Zeeman splitting. As a result, electrons tend to partially occupy the spin-majority LL for any $\lambda>0$, leading to a compressible state. At intermediate $\kappa$, the incompressible state is favored at small $\lambda$, while increasing $\lambda$ drives a transition to the compressible state due to the growing Zeeman energy. In contrast, for sufficiently large $\kappa$, the exchange energy becomes strong enough to stabilize the incompressible state over the compressible one across the entire range of $\lambda$, and no transition to the compressible state occurs.

As the Zeeman coupling increases, the compressible phase expands at the expense of the incompressible phase. We estimate the experimentally relevant value of $\kappa$, marked by the red star on the $x$-axis. For a twist angle $\theta = 3.5^\circ$ and dielectric constant $\epsilon = 10$, the estimated value is $\kappa \approx 2.6$. In this regime, the system lies close to the phase boundary between the compressible and incompressible phases, depending on the strength of the Zeeman coupling. This proximity signals the fragility of the incompressible state along this Středa branch. Our result is consistent with experimental reports~\cite{Cai2023,Zeng2023,Park2023,PhysRevX.13.031037}, where the gap is absent or strongly suppressed.

We summarize the energetic contributions that favor the incompressible state along both St\v{r}eda branches in Table~\ref{tab:energy_competition}.



\section{Spin-flip exciton along St\v{r}eda line Toward charge neutrality} \label{sec:spin-flip}



Since the Zeeman energy does not favor the incompressible Chern-insulator state $|\Psi_{IC}^T\rangle$ along the St\v{r}eda line toward charge neutrality and it increases linearly with $\lambda$, one may suspect that the spin-flip (neutral) excitations could destabilize the incompressible state at sufficiently large $\lambda$. As we show below using time-dependent Hartree–Fock calculation, this indeed occurs.
Because spin-polarization along the $z$-direction  $S_z$ commutes with the Hamiltonian (see Eq.~\eqref{eq:many_body_H}), different $S_z$ sectors decouple. The spin-polarization of the incompressible Chern insulator $|\Psi_{IC}^T\rangle$ is
 $S_z = -\hbar N_\downarrow/2$. We focus on a single spin-reversal excitation in which one particle flips from spin-$\downarrow$ to spin-$\uparrow$. The spin-$\downarrow$ must come from the occupied Landau level but the spin-$\uparrow$ electron can occupy any Landau level. These single spin-reversal excitations are
 described by acting spin-reversal operator $\hat{O}$ on the Hartree-Fock groundstate,
\begin{figure*}
    \centering
\includegraphics[width=1\linewidth]{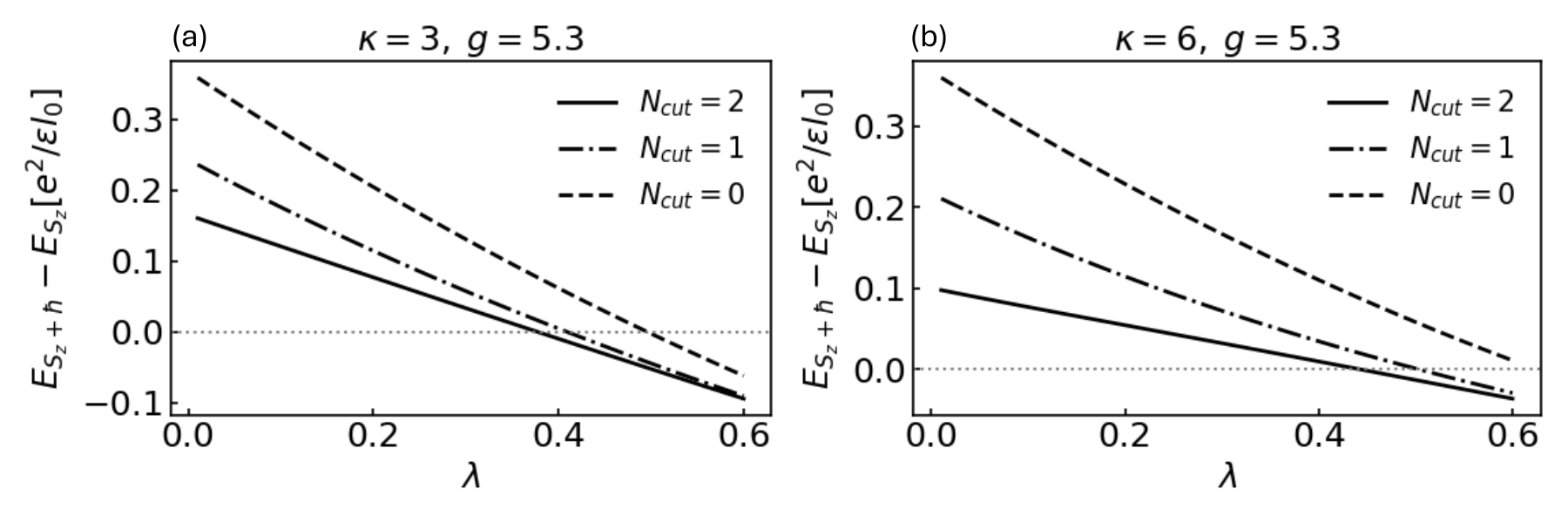}
\caption{Spin-flip exciton excitation energy (the lowest eigenvalue of the RPA equation) of the incompressible Chern insulator along the St\v{r}eda branch pointing toward charge neutrality, shown as a function of $\lambda$ for fixed $\kappa$ and $g$. Including Landau-level mixing (increasing $N_{\mathrm{cut}}$) shifts the instability to smaller values of $\lambda$.}
    \label{fig:excitation_LLs}
\end{figure*}

\begin{align}
    \hat{O}|\Psi_{IC}^T\rangle &= \sum_{m,m'}\sum_{n=0}^{N_{cut}}
    \chi_{n m,0 m'}\, 
    c^\dagger_{n m \uparrow}\,
    c_{0 m' \downarrow}|\Psi_{IC}^T\rangle\\
    &\equiv \sum_{m,m'}\sum_{n=0}^{N_{cut}}
    \chi_{n m,0 m'}\, |(nm)_p\uparrow,(0m')_h\downarrow\rangle
\end{align} 
 The notation $|a_p\uparrow,b_h\downarrow\rangle=c^\dagger_{a_p \uparrow}\,c_{b_h\downarrow}|\Psi_{IC}^T\rangle$ denotes a particle–hole excitation created by flipping one spin. The spin-flip process creates a particle with quantum numbers $a=(n,m)$ in the spin-$\uparrow$ Landau level and leaves behind a hole with quantum numbers $b=(0,m')$ in the spin-$\downarrow$ Landau level. In this notation, the quantum number $m$ for both spin correspond to states with the same chirality.  
 The resulting state is therefore a two-body particle–hole excitation. The coefficients $\chi_{n m,0 m'}$ are obtained by diagonalizing the Hamiltonian in the $|a_p\uparrow,b_h\downarrow\rangle$ using the equation-of-motion method (see Appendix) which leads to the standard random-phase-approximation (RPA) matrix equation
 \begin{equation}
\begin{pmatrix}
A & B \\
-B^* & -A^*
\end{pmatrix},
\end{equation}
Here the $A$ block describes forward-scattering processes, while the $B$ block corresponds to backward amplitudes. In the present case, because $\hat{O}^\dagger|\Psi_{IC}^t\rangle=0$, the backward amplitudes are absent, $B=0$. The RPA matrix therefore reduces to the $A$ block alone. Furthermore, since there is no direct scattering channel, the matrix elements take the following form:
\begin{align}\label{eq:RPA_matrix_ph}
&A_{(0 m_1,\; n m_2),(0 m_1', n' m_2')}
=
(\epsilon_{n\uparrow}-\epsilon_{0\downarrow})
\delta_{n n'}\delta_{m_2 m_2'}\delta_{m_1 m_1'} \notag \\
&-
\langle (n' m_2')_p \uparrow,(0 m_1)_h\downarrow | V|(n m_2)_p\uparrow,(0 m_1')_h\downarrow  \rangle.
\end{align}
The first term represents the Hartree-Fock quasiparticle energy:
\begin{align}\label{eq:single_particle_HF}
    \epsilon_{n\uparrow} &= (1+\lambda)(n+\frac{1}{2})-0.163 \, g\,\lambda \\
    \epsilon_{0\downarrow} &= \frac{1-\lambda}{2}-\kappa \sqrt{\frac{\pi}{2}}\sqrt{1-\lambda}+0.163\,g\,\lambda
\end{align}
Because there is only a single spin-up electron, $\epsilon_{n\uparrow}$ does not have any exchange energy.

The second term in $A$ represents the exchange scattering terms of the electron-hole pair.
When the electron and hole experience the same magnetic field \cite{xu2025localizedexcitonslandaulevelmixing,kwan2021exciton,kwan2022excitonic,kwan2025textured,sodemann2024halperin}, this matrix element can be simplified by transforming the RPA equation into the center-of-mass and relative guiding-center coordinates, as shown explicitly in our previous work \cite{xu2025localizedexcitonslandaulevelmixing}. Because the Coulomb interaction depends only on the relative coordinate, the RPA matrix becomes block diagonal in the guiding-center index. The resulting eigenvalues $A_m$ are then highly degenerate with respect to the center-of-mass coordinate, forming a bosonic analog of Landau levels for the localzied spin excitons \cite{kwan2021exciton,kwan2022excitonic,kwan2025textured,sodemann2024halperin,xu2025localizedexcitonslandaulevelmixing}. In the present case, however, this simplification is not possible because the electron and hole experience different magnetic fields and therefore do not even share the same Landau-level degeneracy. As a result, the standard center-of-mass transformation no longer block-diagonalizes the interaction.


  
Our key insight for solving this problem is to introduce a new basis, which we refer to as the \textit{center-of-charge} basis. In this basis the Coulomb interaction matrix elements become block diagonal, even though the electron and hole experience different magnetic fields. We discuss in detail this center-of-charge basis in the next section where we talk about two-electron problem with arbitrary different magnetic fields. For now, we only mentioned that the basis transformation between individual guiding center coordinates to the center-of-charge  $\ket{J}_{coc}$ and relative basis $\ket{j}_{r}$ are described in Eq.~\eqref{eq:basis_trans_appendix}.

In the present problem, we find that it is sufficient to retain the spin-$\uparrow$ electron states up to the lowest three Landau levels ($N_{cut}=2$); the low-lying excitation spectrum is already well converged at this level. The basis states we consider are,
\begin{align}\label{eq:RPA_basis}
\big(\,
|0_p\!\uparrow,0_h\!\downarrow\rangle |j_0\rangle_r,\;
|1_p\!\uparrow,0_h\!\downarrow\rangle |j_1\rangle_r,\;
|2_p\!\uparrow,0_h\!\downarrow\rangle |j_2\rangle_r \,
\big),
\end{align}
Here, the notation $|i_p\uparrow,0_h\downarrow\rangle$ specifies the Landau level indices of the particle–hole pair and $|j_n\rangle_r$ is the relative guiding-center state.  In this truncated Hilbert space, the RPA matrix reduces to a $3\times3$ problem 
\begin{align}
    A = \begin{pmatrix}
        A_{00} & A_{01} & A_{02} \\
        A_{10} & A_{11} & A_{12} \\
        A_{20} & A_{21} & A_{22}
    \end{pmatrix},
\end{align}
where $A_{nn'}$ are given by,
\begin{align}\label{eq:RPA_matrix_elements}
    A_{nn'} =& \,(\epsilon_{n\uparrow}-\epsilon_{0\downarrow})\delta_{nn'} \nonumber \\
    &-\,{}_r\langle j_n| \langle n \uparrow,0_h\downarrow| V |n'\uparrow,0_h\downarrow\rangle|j_{n'}\rangle_r
\end{align}
The explicit expression is given in Sec.~\ref{sec:RPA_matrix}. Because the total angular momentum is conserved, the lowest Landau level exciton mode ($A_{00}$) can couple to higher Landau-level excitons ($A_{11}$ and $A_{22}$) only if,
\begin{align}
   j_0=j_1-1=j_2-2
\end{align}
This selection rule follows from the conservation of the center-of-charge angular momentum during the exchange-scattering process appearing in $A_{nn'}$.

Fig.~\ref{fig:excitation_LLs} shows the lowest eigenvalue of the RPA matrix, corresponding to the sector with $j_0=0$. We denote this excitation energy by $E_{S_z+\hbar}-E_{S_z}$. To understand how this mode evolves with $\lambda$, it is useful to examine the diagonal element $A_{00}$, which describes the spin-flip exictation within the lowest Landau level ($n=0$):
\begin{align}
A_{00} = \lambda + \kappa \sqrt{1-\lambda}\sqrt{\frac{\pi}{2}}
- \kappa \sqrt{1-\lambda^2}\, V_{j_0}- 0.326\, g \lambda.
\end{align}
Here $V_{j_0}$ is the usual Haldane pseudopotential in the lowest Landau level (see Eq.~\eqref{eq:Haldane_n0_m0}). 
The first term in $A_{00}$ represents the cyclotron-energy difference between the majority-spin and minority-spin lowest Landau levels. The second term corresponds to the self-energy cost of removing an electron from a completely filled lowest Landau level. A finite external magnetic field reduces this exchange self-energy by a factor $\sqrt{1-\lambda}$, because the effective magnetic field experienced by the majority spin decreases from $B_{\mathrm{sky}}$ to $B_{\mathrm{sky}}-B$.
The third term in $A_{00}$ describes exchange scattering between the particle and the hole. Physically, it can be viewed as the binding energy of the exciton arising from the Coulomb attraction between the particle and the hole. This contribution is reduced by a factor $\sqrt{1-\lambda^2}$, due to modification of Coulomb matrix elements in the  center-of-charge basis. 
The last term in $A_{00}$ is the Zeeman-energy and note that it favors the creation of spin-flip excitation because the majority spin aligns the direction opposite to the external magnetic field.

Focusing on the interaction contributions (the terms proportional to $\kappa$) in $A_{00}$, we note that the interaction energy cost of creating a spin exciton decreases with $\lambda$. This is because $\sqrt{1-\lambda^2} > \sqrt{1-\lambda}$ for $0<\lambda<1$, so the particle–hole binding term decreases more slowly than the self-energy term.
Furthermore, since $g > 1/0.326$, the Zeeman splitting is more important than the orbital (cyclotron) splitting. As a result, the single-particle contributions (the terms in $A_{00}$ not proportional to $\kappa$) also favor the creation of a spin exciton.
Taken together, these effects imply that the energy required to create a spin exciton decreases monotonically with increasing $\lambda$. This behavior is shown in the dashed line ( $N_{cut}=0$) in Fig.~\ref{fig:excitation_LLs}a) and b).


When additional Landau levels are included ($N_{\mathrm{cut}}=1,2$), level repulsion develops between the $n=0$ and $n=1$ exciton branches, which shifts the onset of the instability to smaller values of $\lambda$, as shown in Fig.~\ref{fig:excitation_LLs}. This level repulsion indicates that Landau-level mixing favors the formation of spin excitons \cite{xu2025localizedexcitonslandaulevelmixing,PhysRevLett.134.046501,PhysRevB.111.125127}.
We have also checked convergence by including additional Landau levels and find that the eigenvalues change negligibly for realistic interaction strengths $\kappa$ in the range $2$–$8$. This justifies our Landau-level cut-off  $N_{\mathrm{cut}}=2$.

In the phase diagram shown in Fig.~\ref{fig:main_result}, the green region marks correspond to the $\kappa-\lambda$ space where the lowest spin-flip excitation energy of the incompressible state becomes negative. Within this region, the incompressible state is unstable. However, the present calculation cannot determine the nature of the new ground state that emerges once this instability occurs, nor whether the transition is first order or continuous.

\section{Two-Body Problem and the Haldane pseudopotential}
\label{sec:two_body_opposite}
\begin{figure}
    \centering
    \includegraphics[width=1\linewidth]{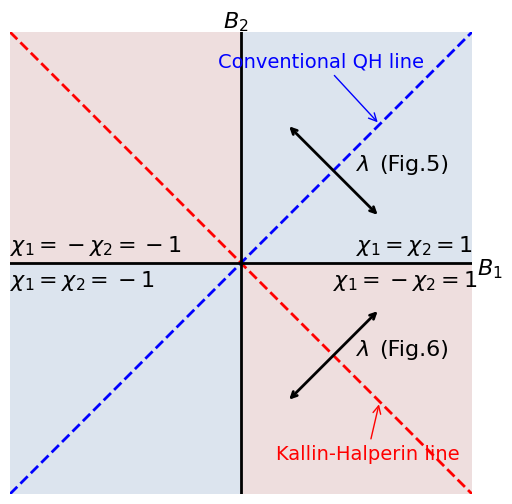}
    \caption{Schematic diagram illustrating the magnetic-field parameter space 
for the two-body problem. $\mathbf{B}_1$ and $\mathbf{B}_2$ denote the magnetic 
fields acting on particles 1 and 2. The blue region indicates 
the sector in which the two fields are aligned, whereas the red region 
corresponds to anti-aligned field orientations. The red dashed line marks the 
Kallin--Halperin line, $B_1=-B_2$, where the two-body problem 
leads to a propagating exciton \cite{PhysRevB.30.5655}. The blue dashed line corresponds to the 
conventional quantum Hall regime $B_1=B_2$. Our results show that all eigenstates 
are localized except along the Kallin--Halperin line. The black arrows indicate 
the direction in the $B_1$--$B_2$ parameter space along which the generalized 
Haldane pseudopotentials evolve, as shown in Figs.~5 and~6
}
    \label{fig:magnetic_field_phase_space}
\end{figure}

In this section, we solve the two-body problem where the two particles experience two distinct uniform magnetic fields in the presence of an isotropic interaction using the center-of-charge basis.
The two-body Hamiltonian consists of two terms:
\begin{align}
H^{2b}&= h_0 +V(|\mathbf{r}_1-\mathbf{r}_2|),\;
h_0 =\sum_{i=1,2}\frac{(\boldsymbol{\pi_i})^2}{2m}\label{eq:free_two_H}
\end{align}
where $h_0$ is the free Hamiltonian for the two particles and $\boldsymbol{\pi}_i =
\mathbf p_i+e \mathbf A_{i}(\mathbf r_i)$ is the kinetic momentum. $V(|\mathbf r_1 - \mathbf r_2|)$ is an isotropic interaction between them. The vector potentials correspond to two distinct uniform magnetic fields $\nabla \times \mathbf A_i=B_i\hat z$ for $i=1,2$. We parameterize the magnetic fields as
\begin{equation}
    B_1=\chi_1(1+\lambda) B\;\;,\;\; B_2=\chi_2(1-\lambda) B
\end{equation}
The parameters $\chi_1$ and $\chi_2$ select the quadrant in Fig.~\ref{fig:magnetic_field_phase_space}. The case $\lambda=0$ with $\chi_1=\chi_2=\pm1$ corresponds to the blue line in Fig.~\ref{fig:magnetic_field_phase_space}. This is the familiar situation in which two electrons move in the same magnetic field; in this case the spectrum is parameterized by the Haldane pseudopotentials. 
For $\lambda=0$ with $\chi_1=-\chi_2=\pm1$ (fourth and second quadrant), the parameters lie on the red dashed line in Fig.~\ref{fig:magnetic_field_phase_space}. This problem was studied by Kallin and Halperin \cite{PhysRevB.30.5655}, who showed that the spectrum consists of propagating waves whose two-dimensional wavevector $\mathbf q$ is related to the dipole separation between the electron and hole, $\mathbf d = \hat z \times \mathbf q l_B^2 $.
In this work we extend the problem to the full $(B_1,B_2)$ parameter diagram. 

For particles moving in a magnetic field, it is convenient to decompose their physical position into cyclotron coordinates, denoted by $\boldsymbol{\eta}$, and guiding-center coordinates, denoted by $\mathbf R$, as
\begin{align}\label{eq:position_decomposition}
    \mathbf r_i =  \boldsymbol \eta_i + \mathbf R_i , \quad \boldsymbol\eta_i = \chi_i l_i^2 \boldsymbol \pi_i \times \hat z,
\end{align}
where the magnetic length is $l_i = \sqrt{\hbar/(e|B_i|)}$. The kinetic term only depends on the cyclotron coordinates while the interaction depends on both.  The commutation relations of the guiding-center and cyclotron coordinates are
\begin{align}
    &[R_{i a},R_{j b}] = -i\chi_i l_i^2\epsilon_{ab}\delta_{ij}\,, \,
    [\eta_{i a},\eta_{j b}] = +i\chi_i\,l_i^2\,\epsilon_{ab}\delta_{ij}, \label{eq:single_particle_ladder}\\ 
    &[R_{i a},\eta_{j b}] = 0.
\end{align}

Since these variables form canonical conjugate pairs, it is convenient to introduce ladder operators.  The cyclotron coordinates $\boldsymbol{\eta}$ correspond to Landau-level indices $n$, while the guiding-center coordinates $\mathbf R$ correspond to guiding-center quantum numbers $m$. The total Hilbert space is therefore $|n_1,m_1\rangle\otimes |n_2,m_2\rangle$. Along the blue line, it is customary to transform to center-of-mass and relative coordinates for both the cyclotron coordinates $\boldsymbol{\eta}$ and the guiding-center coordinates $\mathbf R$. However, away from the blue line the center-of-mass and relative coordinates no longer commute with each other, and it is unclear what is the most  convenient basis to solve the problem.

Our key observation is that there exists a basis called \emph{center-of-charge} basis that block diagonalizes the interaction. 
We label the center-of-charge and relative state as $|NM\rangle_{c+}|nm\rangle_{r+}$ (or sometimes just $|NM\rangle_{c}|nm\rangle_{r}$) for $\chi_1=\chi_2$ and $|NM\rangle_{c-}|nm\rangle_{r-}$ for $\chi_1=-\chi_2$ where the capital letters refer to center-of-charge quantum number and lower case letters refer to relative quantum number.
For any isotropic two body interaction, it satisfies
\begin{align}\label{eq:block_diagonal_V_COC}
    &{}_{\pm c}\langle NM|{}_{\pm r}\langle nm|V|n'm'\rangle_{r\pm}|N'M'\rangle_{c\pm} \notag \\
    &= {}_{\pm r}\langle nm|V|n'm'\rangle_{r\pm} \delta_{NN'}\delta_{MM'}\delta_{n-m,n'-m'}
\end{align}
This retains exactly the same form as the isotropic two body interaction in the center-of-mass and relative basis in conventional quantum Hall and provides a general method for treating interactions when particles experience different magnetic fields. 

The motivation for introducing the center-of-charge basis is that
instead of treating the two electrons in $h_0$ as experiencing different magnetic fields, we can equivalently view them as moving in the same magnetic field but carrying different effective charges,
\begin{align}\label{eq:effective charge}
    q_1 = \chi_1 (1+\lambda)e, \qquad q_2 = \chi_2 (1-\lambda)e,
\end{align}
We then define the center-of-charge and relative guiding-center coordinates as
\begin{align}
\mathbf r_c &= \frac{q_1 \mathbf r_1+q_2 \mathbf r_2}{q_1+q_2} = \boldsymbol{\eta}_c+\boldsymbol{R}_c, \\
\mathbf r_r &= \mathbf r_1 - \mathbf r_2=\boldsymbol{\eta_r}+\boldsymbol{R}_r. \label{eq:rr_def}
\end{align}
where $\boldsymbol{\eta_c}$ and $\boldsymbol{\eta_r}$ denote the center-of-charge and relative coordinates and  $\vec R_c$ and $\vec R_r$ are the guiding center version.
We keep the original definition of relative distance between the interaction only depends on $|\vec r_1-\vec r_2|$.
The corresponding conjugate momenta are given by:
\begin{align}
    \mathbf p_c = \mathbf p_1 + \mathbf p_2, \quad \mathbf p_r = \frac{q_2 \mathbf p_1 - q_1 \mathbf p_2 }{q_1+q_2} \label{eq:Pp_def}
\end{align}
The center-of-charge and relative cyclotron and guiding center coordinates are defined in the same manner.  
The corresponding commutation relations are given by:
\begin{align}\label{eq:commutation_relation_general}
    &[R_{ra},R_{rb}] = -i \frac{\chi_1(1-\lambda)+\chi_2(1+\lambda)}{1-\lambda^2}l_0^2 \epsilon_{ab}, \\
    &[R_{ca},R_{cb}] = -i \frac{l_0^2}{\chi_1(1+\lambda)+\chi_2(1-\lambda)}\epsilon_{ab},\\
    & [R_{ca},R_{rb}] = 0, \;\;\; l_0 = \sqrt{\frac{\hbar}{eB}}
\end{align}
The commutation relations for cyclotron coordinates have the same form up to a minus sign.
Note that the commutation relations become singular along three lines in Fig.~\ref{fig:magnetic_field_phase_space}: $B_1=0$ and $B_2=0$ which correspond to $\lambda=1$, and the Kallin–Halperin line $B_1=-B_2$ which correspond to $\chi_1=-\chi_2$ and $\lambda=0$. At $B_1=0$ or $B_2=0$, the singularity arises because one of the particles does not experience a magnetic field. At $B_1=-B_2$, the eigenstate is a propagating wave and cannot be decoupled into center-of-charge and relative state. 
In the center-of-charge and relative coordinates, the two-body Hamiltonian in Eq.~\eqref{eq:free_two_H} becomes
\begin{align}
   H^{2b}
&= \frac{q_1^2+q_2^2}{(q_1+q_2)^2} \frac{1}{2m}\mathbf \Pi^2+\frac{1}{m}\boldsymbol \pi^2 
    +\frac{q_1-q_2}{q_1+q_2}\frac{\mathbf \Pi\cdot\boldsymbol\pi}{m} +V(|\vec r_r|)\label{eq:COM_H}
\end{align}
where $\mathbf{\Pi}$ and  $\pi$ are the kinetic center-of-charge and relative momenta, its explicit form in the symmetric gauge read as:
\begin{align}
    \mathbf{\Pi}&= \mathbf p_c+(q_1+q_2)\frac{\mathbf B\times \mathbf r_c}{2}
    \\\boldsymbol \pi&= \mathbf p_r + \frac{q_1q_2}{q_1+q_2} \frac{\mathbf B \times \mathbf r_r}{2}
\end{align}
The first term in the Hamiltonian  $H^{2b}$ describes the center-of-charge coordinate as a particle with an 
effective charge $q_1+q_2$ and 
effective mass $\frac{(q_1+q_2)^2m}{q_1^2+q_2^2}$ moving in a magnetic field $\boldsymbol B$.  The second term describes the relative coordinate as a particle with an effective charge $\frac{q_1q_2}{q_1+q_2}$ and effective mass $m/2$ moving in the same magnetic field $\boldsymbol B$. However, the presence of the third term does not decouple the center-of-charge and relative motion. 




%



In many situations we are interested in projecting the problem onto a given Landau level.  In that case the kinetic energy is fixed, and the projected Hamiltonian takes the form
\begin{align}
H_{n}^{2b}(\mathbf R_1,\mathbf R_2)&=
\hbar(\omega_{1}+\omega_{2})\left(n+\tfrac{1}{2}\right)
+
V_{n}(|\mathbf R_1-\mathbf R_2|), \\
V_{n}(|\mathbf R_1-\mathbf R_2|) &=  \int \frac{d^2q}{(2\pi)^2}V(|q|) F_{n}(\frac{ql_0}{\sqrt{1+\lambda}})F_{n}(-\frac{ql_0}{\sqrt{1-\lambda}}) \notag \\
&\times e^{i\mathbf q \cdot (\mathbf R_{1}-\mathbf R_{2})}
\end{align}
where $\omega_i$ is the cyclotron frequency and 
$F_n(x) =  \mathrm L_n\left(\frac{x^2}{2}\right) e^{-\frac{x^2}{4}}
$ is Landau level form factor.
Therefore, the only remaining term to diagonalize is the projected interaction, whose eigenvalues can be interpreted as \emph{generalized Haldane pseudopotentials}. Below, we study their evolution across the $(B_1,B_2)$ plane, away from the conventional quantum Hall line and the Kallin–Halperin line, by moving along the perpendicular direction as illustrated in Fig.~\ref{fig:magnetic_field_phase_space}. Because the sign of the commutation relations in Eq.~\eqref{eq:commutation_relation_general} changes between different quadrants of the $B_1$–$B_2$ plane, it is convenient to treat the cases $\chi_1=\chi_2$ and $\chi_1=-\chi_2$ separately in the following.





\subsection{$\chi_1=\chi_2=1$}\label{sec:chi_1=chi_2}
We set $\chi_1=\chi_2=1$ in Eq.~\eqref{eq:commutation_relation_general}, corresponding to the first quadrant in Fig.~\ref{fig:magnetic_field_phase_space}. The same result holds for the third quadrant. The resulting commutator relationship lead to the following definition of ladder operators for the center-of-charge and relative coordinates,
\begin{align}\label{eq:ladder_operators_def}
& B_c =\sqrt{\frac{1}{l_0^2}} \left( R_{cx} - i R_{cy} \right), \quad [B_c,B_c^\dagger]=1 ,\\
&b_{r} = \frac{1}{2}\sqrt{\frac{1-\lambda^2}{l_0^2}}\left( d_x-id_y \right) ,\quad [b_r,b_r^\dagger]=1 ,\\
& [b_r,B_c] = [b_r,B_c^\dagger]=0 ,
\end{align}
where $ l_0 = \sqrt{\hbar/(e B)}$. The corresponding center-of-charge and relative state are defined as
\begin{align}\label{eq:COC--relative state same direction}
|M\rangle_{c+}|m\rangle_{r+} =
\frac{(B_c^\dagger)^M (b_r^\dagger)^m}{\sqrt{M!m!}}|0\rangle_{+} ,
\end{align}
where $|0\rangle_{+}$ is the vacuum annihilated by $B_c$ and $b_r$. The corresponding center-of-charge and relative wavefunctions can be obtained from the center-of-mass and relative wavefunctions in the usual quantum Hall problem by rescaling the magnetic length. The basis transformation from the center-of-charge basis to the individual basis is presented in Appendix.~\ref{sec:basis_trans}.

Since the isotropic interaction is diagonal in the center-of-charge and relative basis,
the generalized Haldane pseudopotentials are given by:
\begin{align}
  V^{+}_{n,m}&\equiv {}_{+}\langle n,n; M,m | V | n,n; M,m \rangle_{+} , \\&= \int \frac{d^2q}{(2\pi)^2}V(|\mathbf q|)F_{n}\left(\frac{ql_0}{\sqrt{1+\lambda}}\right) F_{n}\left(-\frac{ql_0}{\sqrt{1-\lambda}}\right) \notag\\ &\times L_m\left(\frac{q^2l^2_0}{1-\lambda^2}\right) e^{-\frac{l_0^2  q^2}{2(1-\lambda^2)}}\label{eq:analyticall V_1}
\end{align}
where $| n,n; M,m \rangle_{+}\equiv|M\rangle_{c+}|m\rangle_{r+} |n,n\rangle$ and we have used the relation below to obtain the above expression,
\begin{align}
        {}_{c}\langle M|{}_{r}\langle m|e^{i\mathbf q\cdot (\mathbf R_1-\mathbf R_2)}|M\rangle_{c}|m\rangle_{r} =L_m(\frac{q^2l^2_0}{1-\lambda^2})e^{-\frac{l_0^2  q^2}{2(1-\lambda^2)}} 
\end{align}
The $+$ superscript in the last line is to remind us that $\text{sgn}(B_1)=\text{sgn}(B_2)$.
We can obtain simple analytic expressions in several cases for Coulomb interaction. For $n=0$, 
\begin{align}\label{eq:Haldane_n0_m0}
    V^{+}_{0,m} = \frac{e^2}{\epsilon l_0} \sqrt{1-\lambda^2} V_m
\end{align}
where $V_m = \Gamma(m+\tfrac{1}{2})/(2\,\Gamma(m+1))$ is the usual Haldane pseudopotential in the lowest Landau level.  For $n=1$, the pseudopotentials is not a simple rescaling but takes the following form
\begin{align}
    V^{+}_{1,m}  = \frac{e^2}{\epsilon l_0}\left( V_{1,m}(1-\lambda^2)^{\frac{3}{2}}+\lambda^2 {}_r \langle 1;m|V_C|1;m\rangle_r \right)\label{eq:n1_gpseudopotential}
\end{align}
where $V_{1,m}$ is the Haldane pseudopotential in the $n=1$ Landau level and ${}_r \langle n;m|V_C|n;m\rangle_r$ is the Coulomb matrix elements in the relative  basis defined in Eq.~\eqref{eq:Coulomb_relative}. 
\begin{figure}
    \centering
    \includegraphics[width=1\linewidth]{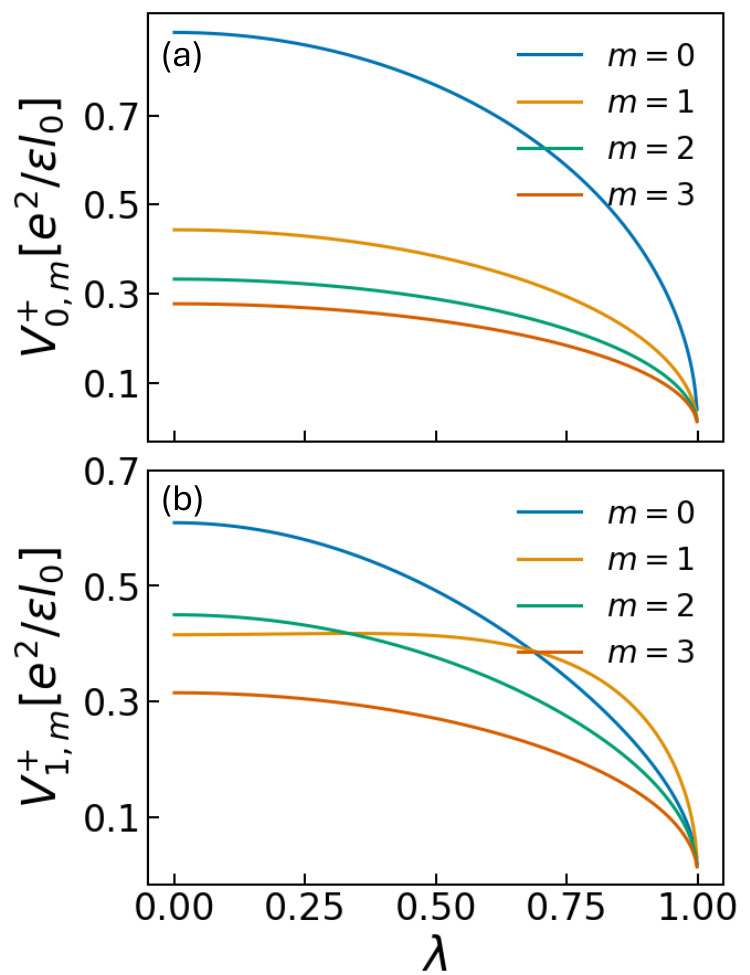}
\caption{Generalized Haldane pseudopotentials projected onto the $n=0$ (a) and $n=1$ (b) Landau levels as a function of $\lambda$. Increasing $\lambda$ corresponds to moving perpendicular to the blue dashed line that marks the conventional quantum Hall limit in Fig.~\ref{fig:magnetic_field_phase_space}.}
    \label{fig:eigenvalues_same_direction}
\end{figure}

Fig.~\ref{fig:eigenvalues_same_direction} shows the generalized Haldane pseudopotentials 
$V_{0,m}^{+}$ [panel (a)] and $V_{1,m}^{+}$ [panel (b)] as functions of the  
$\lambda$ for different components $m$.
In the $n = 0$ Landau level, the $\lambda$ dependence is trivial, as the  
pseudopotential simply scales as $\sqrt{1 - \lambda^2}$.  Consequently, all components  decrease monotonically with increasing $\lambda$ without crossing.


In contrast, the $\lambda$ dependence of $V_{1,m}^{+}$ in Fig.~\ref{fig:eigenvalues_same_direction}(b) exhibits a level crossing. For $\lambda \ll 1$, the first term in Eq.~\eqref{eq:n1_gpseudopotential} dominates over the second term. In this regime, the generalized Haldane pseudopotentials follow the conventional ordering of the $n=1$ Landau level, $V_{1,m=0}>V_{1,m=2}>V_{1,m=1}>V_{1,m=3}....$.

However, as $\lambda \rightarrow 1$, approaching the limit where one of the electrons effectively experiences no magnetic field, the second term in $V_{1,m}^{+}$ becomes dominant. In this limit, the matrix element ${}_r\langle 1;m|V_C|1;m\rangle_r$ is largest for $m=1$, corresponding to physical relative angular momentum $(m-n)=0$, where the overlap between the two-particle wavefunctions is maximal. As a result, the pseudopotentials acquire an anomalous ordering, $V_{1,m=1}>V_{1,m=0}>V_{1,m=2}>V_{1,m=3}....$. At intermediate values of $\lambda$, the pseudopotentials follow a sequence similar to that of the $n=0$ Landau level, $V_{1,m=0}>V_{1,m=1}>V_{1,m=2}>V_{1,m=3}....$.




\subsection{$\chi_1=-\chi_2=1$}

In this section, we set $\chi_1=-\chi_2=1$ in Eq.~\eqref{eq:commutation_relation_general}, which correspond to the fourth quadrant. The same result holds for the second quadrant.
The corresponding center-of-charge and relative ladder operators in this case are defined as: 
:
\begin{align} 
&\bar B_c = \frac{\sqrt{|\lambda|}}{l_0}(R_{c x}-i R_{c y}), \quad [\bar B,\bar B^\dagger]=1,  \label{eq:commutation_1}\\
    &\bar b_r = \frac{1}{2}\sqrt{\frac{1-\lambda^2}{|\lambda| l_0^2}}(d_x+i d_y), \quad [\bar b,\bar b^\dagger]=1,\label{eq:commutation_2} \\
    &[\bar b,\bar B]=[\bar b,\bar B^\dagger]=0.
     \label{eq:commutation_3}
\end{align}
Accordingly, the center-of-charge and relative state can be defined as: 
\begin{align}\label{eq:COC-r-state-opposite}
    |M,m\rangle_{-} = \frac{(\bar B^\dagger)^M (\bar b^\dagger)^m}{\sqrt{M!m!}}|0\rangle_{-}
\end{align}
Here the $|0\rangle_{-}$ is the vacuum state that is annihilated by $\bar B$ and $\bar b$. Interesting, this vacuum is annihilated by a superposition of individual lowering and raising ladder operators, see Eq~\eqref{eq:opposite_vacuum}.

Due to the commutation relations described in Eq.~\ref{eq:commutation_1}-\ref{eq:commutation_3}, 
the eigenvalues become discrete (away from the Kallin–Halperin line) and resemble the form of the Haldane pseudopotentials:
\begin{align}
    V_{n,m}^{-}&=\;_{-}\langle n,n;M,m||V|n,n;M,m\rangle_{-}
      \notag \\
    &=\int \frac{d^2q}{(2\pi)^2}V(|\mathbf q|)F_{n}(\frac{ql_0}{\sqrt{1+\lambda}}) F_{n}(-\frac{ql_0}{\sqrt{1-\lambda}})
  \notag \\
   &\times L_m(q^2l^2_0\frac{|\lambda|}{1-\lambda^2})e^{-\frac{|\lambda| l_0^2}{2(1-\lambda^2)} q^2}.
\end{align}
where $|n,n;M,m\rangle_{-}\equiv|M\rangle_{c-}|m\rangle_{r-} |n,n\rangle$ and we used the relation,
\begin{align}
    &{}_{-c}\langle M|{}_{-r}\langle m|e^{i\mathbf q\cdot (\mathbf R_1-\mathbf R_2)}|M\rangle_{c-}|m\rangle_{r-} \notag \\
    &=L_m(q^2l^2_0\frac{|\lambda|}{1-\lambda^2})e^{-\frac{|\lambda| l_0^2}{2(1-\lambda^2)} q^2}
\end{align}

We now discuss the generalized Haldane pseudopotential for Coulomb potential projected into the $n=0$ Landau level  and  $n=1$ Landau level . In the $n=0$ Landau level, the eigenvalues are given by
\begin{align}
    V_{0,m}^{-}  = \frac{e^2}{\epsilon l_0} \sqrt{\frac{\pi}{2}}\sqrt{1-|\lambda|}\,\ {}_2F_1\!\left(-m,\frac{1}{2};1;\frac{2|\lambda|}{1+|\lambda|}\right). \label{eq:eigenvalue_n0}
\end{align}
where $_2F_1$ is Gauss hypergeometric function. While for the $n=1$ Landau level, the explicit form is very long and not illuminating and not shown here. 


\begin{figure}
    \centering
    \includegraphics[width=1\linewidth]{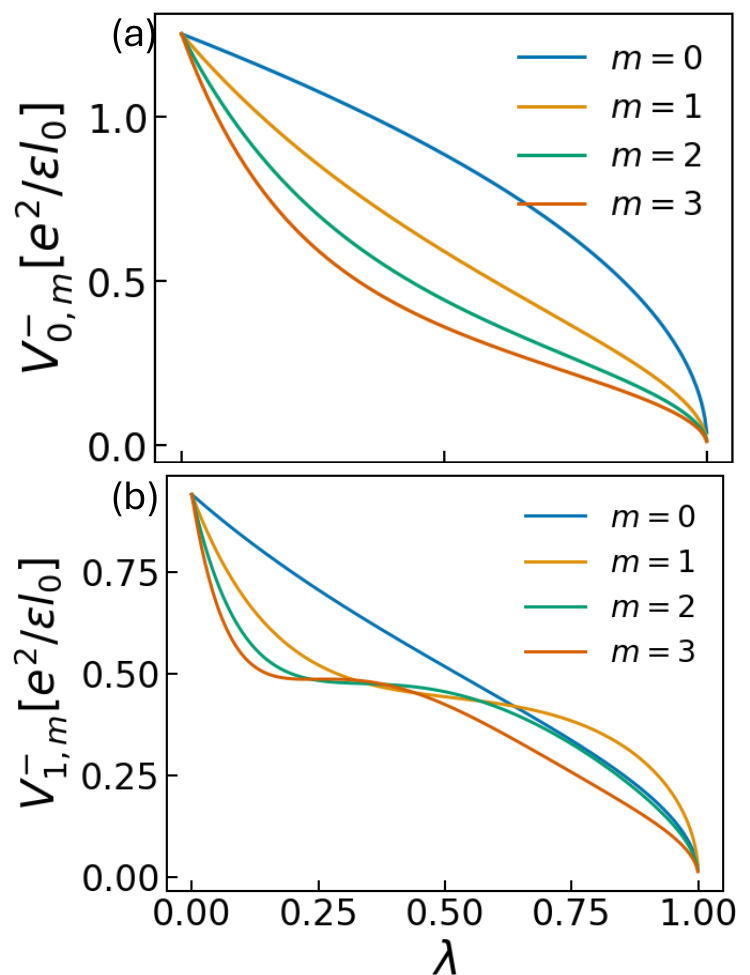}
    \caption{The generalized Haldane pseudopotentials vs $\lambda$ projected onto the $n=0$ (a) and $n=1$ (b) Landau levels.
    Increasing $\lambda$ corresponds to moving perpendicular to the red dashed line that marks the Kalline-Halperine line in Fig.~\ref{fig:magnetic_field_phase_space}.
    }
    \label{fig:Coulomb_eigenvalues_n0_n1}
\end{figure}

Fig.~\ref{fig:Coulomb_eigenvalues_n0_n1}(a) shows the various components of the eigenvalues $V_{0,m}^{-}$ in the $n=0$ Landau level as functions of $\lambda$. In the limit $\lambda \rightarrow 0$, ${}_2F_1(a,b;z=0)=1$, and the pseudopotential reduces to $V_{0,m}^{-}=\sqrt{\pi/2}\,(e^2/(\epsilon l_0))$. This corresponds to the Coulomb energy of a particle (positive $B$) and a hole (negative $B$) sitting on top of one another. This energy exactly cancels the exchange energy of a completely filled zeroth Landau level, producing the intra-Landau level spin-flip mode discussed in the Kallin–Halperin analysis \cite{PhysRevB.30.5655}. For $\lambda \ll 1$, the differences between different pseudopotentials are small.
As $\lambda$ increases further, these differences shrink and eventually collapse to zero. This occurs because one particle effectively feels no magnetic field and its eigenstate is a plane wave, so the two-particle state can no longer be localized.



Fig.~\ref{fig:Coulomb_eigenvalues_n0_n1}(b) shows the $n=1$ Landau level projected Haldane pseudopotential $V_{1,m}^-$.
It has the same behavior in the $n=0$ Landau level for the limits $\lambda \rightarrow 0$ and $\lambda \rightarrow 1$.
In addition to decreasing with $\lambda$, the eigenvalues also exhibit complicated level crossings as a function of $\lambda$.
For small $\lambda$, the ordering of the eigenvalue components follows the natural sequence of increasing $m$. For $\lambda \sim 0.25\text{--}0.5$, the $m>0$ pseudopotentials become nearly degenerate.

To better understand these results, we compute the spatial structure of the two-particle states by evaluating the expectation values of the relative and physical center-of-mass  coordinates in the $n=0$ Landau level:
\begin{align} {}_{-}\langle M,m|| z_1- z_2|^2|M,m\rangle_{-} &= \frac{4\lambda}{1-\lambda^2}l_0^2(m+1),\label{eq:relative_z}
\\ {}_{-}\langle M,m||\frac{z_1+z_2}{2}|^2|M,m\rangle_{-} &= \frac{M l_0^2}{\lambda} +\frac{(m+1)l_0^2}{2\lambda^2(1-\lambda^2)} \notag \\ &+\frac{l_0^2}{2(1+\lambda)}. 
\end{align}
Eq.~\eqref{eq:relative_z} corresponds to the conventional quantum Hall case of $l_0^2(m+1)$ but multiplied by $\frac{4\lambda}{1-\lambda^2}$. In the limit $\lambda\to0$, the relative distance vanishes and the two particles sit on top of each other. 
The dependence on $m$ is similar to that in the conventional quantum Hall problem.

However, the behavior of the COM position is more complicated. The first term resembles the conventional quantum Hall result and depends only on the center-of-charge angular momentum $M$, while the second term depends on the relative angular momentum $m$. This occurs because the physical COM coordinates do not commute with the relative coordinates, and this coupling leads to a non-monotonic dependence of the COM position on $\lambda$.

In the limit $\lambda \rightarrow 0$, the expectation value of the relative separation vanishes, while the center-of-mass coordinate diverges. Physically, this corresponds to the situation in which the two particles experience magnetic fields of equal magnitude but opposite sign. In this limit, the problem reduces to the Kallin–Halperin description of propagating excitons, with wave vector $\vec q = \hat{z}\times\vec{d}/l_B^2 \rightarrow 0$.
For finite $\lambda<1$, both the relative and COM expectation values are finite, indicating that the two-particle state is localized.

\section{Summary}\label{sec:conclusion}

In this work, we investigated the effect of an external magnetic field on a time-reversal-symmetric Hamiltonian consisting of Landau levels with opposite chirality. This problem is partly motivated by recent magnetotransport experiments in twisted MoTe$_2$ \cite{Cai2023,Zeng2023,Park2023,PhysRevX.13.031037}, where incompressible states appear asymmetrically along the St\v{r}eda lines in the density–magnetic-field phase diagram. Using a new basis, which we term the center-of-charge basis, we obtain analytical expressions for the mean-field energies and the spin-flip exciton spectrum, thereby providing a simple physical explanation for the observed magnetotransport features.

Within the Hartree–Fock approximation, we find that the relative stability of compressible and incompressible states is governed by both universal and model-dependent effects. The exchange energy always favors the incompressible state along either St\v{r}eda line. The kinetic (cyclotron) energy favors the incompressible state along the St\v{r}eda line away from charge neutrality. These contributions represent the universal aspects of the energetics. In contrast, the Zeeman energy introduces model-dependent physics. Combining these ingredients for MoTe$_2$ leads to phase diagrams consistent with recent experimental observations \cite{Cai2023,Zeng2023,Park2023,PhysRevX.13.031037}. One interesting future work is focusing on understanding the nature of the spin-flip instability at large magnetic fields.

We also note recent magnetotransport experiments on twisted WSe$_2$ \cite{doi:10.1126/science.adi4728}, which report behavior opposite to that observed in twisted MoTe$_2$. In twisted WSe$_2$, a prominent gap appears along the St\v{r}eda line pointing toward charge neutrality, while no comparable incompressible feature is observed along the St\v{r}eda line pointing away from neutrality. Our theory can account for this situation if the Chern number of the moiré miniband and the sign of the skyrmion-induced magnetic field in the $K$ valley is opposite to that assumed in this work for MoTe$_2$.


A key technical development of this work is the introduction of a new basis, which we term the center-of-charge basis, that allows the two-body problem to be solved even when the two electrons experience different magnetic fields. Using this basis, we solve the two-particle problem and show that the eigenstates are generally localized, except along the Kallin–Halperin line, where propagating excitons emerge. More broadly, the center-of-charge basis introduced here provides a useful framework for treating two-body problems in which particles experience different magnetic fields and may find applications in a wide range of quantum Hall and moiré systems.

\

\textbf{Acknowledgment} We are grateful to Nemin Wei and Inti Sodemann for insightful discussions.

\bibliographystyle{ieeetr}

\bibliography{reference}

\appendix

\pagebreak
\newpage

\appendix

\section{Hartree-Fock Energy per Particle}\label{sec:energy_per_prticles}

In this section, we present the details for obtaining the energy per particle of the compressible state $|\Psi_{\text{CO}}^A\rangle$ along the St\v{r}eda line away from charge neutrality. The corresponding expression along the St\v{r}eda line toward charge neutrality, namely for $|\Psi_{\text{CO}}^T\rangle$, can be obtained in the same manner.

The density matrix of the uniform state $|\Psi_{\text{CO}}^A\rangle$ satisfies Eq.~\eqref{eq:compressible_state}:
\begin{align}
     \langle \Psi_{CO}^A|c^\dagger_{m\uparrow}c_{m'\uparrow}|\Psi_{CO}^A\rangle &= \delta_{mm'}\frac{N_\uparrow-N_\downarrow}{N_{\uparrow} }= \delta_{mm'} \frac{2\lambda}{1+\lambda} \\
     \langle \Psi_{CO}^A|c^\dagger_{m\downarrow}c_{m'\downarrow}|\Psi_{CO}^A\rangle &= \delta_{mm'},\label{eq:compressible_state_condition2}
\end{align}
Since there is no inter-spin correlation, the total energy can be separated into two contributions: the first arises from particles partially occupying the spin-$\uparrow$ Landau level, and the second arises from particles fully occupying the spin-$\downarrow$ Landau level:
\begin{align}
    \varepsilon^{A}_{\text{CO}} = \frac{1}{N_\uparrow} \langle \Psi_{\text{CO}}^A|H|\Psi_{\text{CO}}^A\rangle = \frac{E_\uparrow+E_{\downarrow}}{N_\uparrow}
\end{align}
Here $N_\uparrow$ denotes the number of particles along the St\v{r}eda line away from charge neutrality, see Eq.~\eqref{eq:imbalance degen}. 
$E_\downarrow$ is the total energy of the particles that completely fill the spin-$\downarrow$ Landau level, which is given by
\begin{align}
    E_\downarrow = N_\downarrow \left[\frac{1-\lambda}{2}
    -\frac{\kappa}{2}\sqrt{1-\lambda}\sqrt{\frac{\pi}{2}}+0.163 g \lambda \right].
\end{align}
$E_\uparrow$ is the total energy of the particles that partially occupy the spin-$\uparrow$ Landau level.
Since the compressible state is assumed to be uniform, its Fock self-energy scales with the partial filling factor $\nu = (N_\uparrow-N_\downarrow)/N_\uparrow$. This contribution is given by
\begin{align}
    E_\uparrow = (N_\uparrow-N_\downarrow)\left[\frac{1+\lambda}{2}
    - \frac{\kappa}{2} \sqrt{1+\lambda}\sqrt{\frac{\pi}{2}}\nu -0.163g\lambda \right].
\end{align}
Therefore, the energy per particle of the uniform compressible state is then given by
 \begin{align}
    \varepsilon_{CO}^A 
     &= \frac{1+3\lambda^2}{2(1+\lambda)}
        - \frac{\kappa}{2}\sqrt{\frac{\pi}{2}}\left[
             \frac{4\lambda^2}{(1+\lambda)^{3/2}} + \frac{(1-\lambda)^{3/2}}{1+\lambda}
          \right] \notag\\
          & - 0.163\frac{3\lambda^2-\lambda}{1+\lambda} g.
 \end{align}

\section{Basis transformation}\label{sec:basis_trans}
In this section, we obtain the relation of the center-of-charge and relative basis to the individual basis. To do that, we first obtained the relation between center-of-charge and relative ladder operators with individual ladder operators. The individual ladder operator can be defined from commutation relation in Eq.~\eqref{eq:single_particle_ladder} as the following:
\begin{align}\label{eq:individual_ladder}
    b_i = \frac{1}{\sqrt{2}l_i^2}(R_{i x}-i\chi_i R_{i y}), \quad [b_i,b_i^\dagger]=1.
\end{align}
In the next, we consider $\chi_1=\chi_2=1$ and  $\chi_1=-\chi_2=1$ separately. 

\subsection{$\chi_1=\chi_2=1$}

From Eqs.~\eqref{eq:individual_ladder} and \eqref{eq:ladder_operators_def}, and noting that $\chi_1=\chi_2=1$, we can express the center-of-charge and relative ladder operators in terms of the individual ladder operators as 
\begin{align} \label{eq:COM_INDIVIDUAL_SAME} &B_c = \sqrt{\frac{1+\lambda}{2}} b_1 + \sqrt{\frac{1-\lambda}{2}} b_2 \\ & b_r = \sqrt{\frac{1-\lambda}{2}} b_1 - \sqrt{\frac{1+\lambda}{2}} b_2 \end{align} 
These relations imply that the vacuum in the center-of-charge–relative basis is simply the direct product of the individual vacua.
Substituting Eq.~\eqref{eq:COM_INDIVIDUAL_SAME} into Eq.~\eqref{eq:COC--relative state same direction} and applying the binomial expansion, we obtain the relation between the center-of-charge–relative basis and the individual basis:
\begin{align}\label{eq:individual_to_COC} |J\rangle_c|j\rangle_r &= \sum_{l=0}^{J+j} R^{J,j}_{l}(\lambda)\, |J+j-l\rangle \otimes |l\rangle . 
\end{align}
Here $|m_1\rangle\otimes|m_2\rangle$ denotes the product of individual states. We remind the reader that the first particle experiences a magnetic field $(1+\lambda)B$, while the second particle experiences $(1-\lambda)B$.
The coefficients $R^{J,j}_l(\lambda)$ are given by
\begin{align}\label{eq:gR_trans} R^{J,j}_{l}(\lambda) &= \sum_{k=\max(0, l-J)}^{\min(j, l)} (-1)^k\, \binom{J}{l-k} \binom{j}{k} \sqrt{\frac{(J+j-l)!\,l!}{J!\,j!\,2^{J+j}}} \notag \\ &\times (\sqrt{1+\lambda})^{\,J-l+2k}\, (\sqrt{1-\lambda})^{\,j+l-2k}. 
\end{align} The inverse transformation has the same form. The Landau-level indices can also be transferred between the individual and center-of-charge–relative bases, with the same coefficient structure. This will be useful for evaluating the Coulomb interaction matrix elements in Appendix~\ref{sec:Coulomb matrix elements opposite spins}.

\subsection{$\chi_1=-\chi_2=1$}

Setting $\chi_1=-\chi_2=1$ in Eq.~\eqref{eq:individual_ladder}, one obtains the relation between $\bar B$, $\bar b$ and the individual ladder operators $b_i$ as
\begin{align}\label{eq:bogoliiubov_opeartor_type}
    &\bar b_r = \sqrt{\frac{1-\lambda}{2\lambda}}\,b_1^\dagger - \sqrt{\frac{1+\lambda}{2\lambda}}\,b_2, \\
    &\bar B_c =  \sqrt{\frac{1+\lambda}{2\lambda}}\,b_1 - \sqrt{\frac{1-\lambda}{2\lambda}}\,b_2^\dagger.
\end{align}
These relations take the form of a Bogoliubov transformation between the center-of-charge and the individual ladder operators. As a result, the two-body vacuum in the center-of-charge–relative basis is no longer a simple direct product of the individual vacua, but instead becomes a two-mode squeezed state, denoted by $|0\rangle_{-}$ and given by
\begin{align}\label{eq:opposite_vacuum}
|0\rangle_{-}
    & = \sqrt{\frac{2\lambda}{1+\lambda}}
       \sum_{m=0}^{\infty}
       \left(\sqrt{\frac{1-\lambda}{1+\lambda}}\right)^{m}
       |m\rangle\otimes |m\rangle.
\end{align}
We remind the reader that the first particle experiences a magnetic field $(1+\lambda)B$, while the second particle experiences $-(1-\lambda)B$. As a result, the vacuum itself is entangled, and the degree of entanglement decreases monotonically as $\lambda$ increases, diverging in the limit $\lambda \rightarrow 0$. In this limit, the superposition weights become identical across all guiding-center states, indicating a translationally invariant state that corresponds to the $\vec q=0$ Kallin–Halperin state~\cite{PhysRevB.30.5655}. In contrast, as $\lambda \rightarrow 1$, the vacuum reduces to a simple direct product of the individual vacua.

\section{Coulomb Interaction Matrix Elements}\label{sec:Coulomb matrix elements opposite spins}

In this appendix, we present the details for the Coulomb interaction matrix elements in Eq.~\eqref{eq:many_body_H}. 

The Coulomb matrix elements are defined as:

\begin{align}\label{eq:Coulomb_matrix_elements}
    V_{1,2';3,4'}^{\sigma\sigma'} = \langle n_1m_1\sigma,n_2'm_2'\sigma'|V_C^{\sigma\sigma'}| n_3m_3\sigma,n_4'm_4'\sigma'\rangle 
\end{align}

where "1" is a shorthand for $n_1m_1$. The $'$ on the quantum number indicates magnetic field direction of the state.

For the Coulomb interaction between same spin species $\sigma'=\sigma$, since the particles experience the same direction of magnetic fields,  one can transfer the two body state from individual basis to center-of-charge basis with Eq.~\eqref{eq:individual_to_COC}, 
\begin{align}\label{eq:basis_trans_appendix}
    |n_1m_1,n_2m_2\rangle &= \sum_{u,l=0}^{n_1+n_2, m_1+m_2} R^{n_1+n_2}_{u}(\lambda) R^{m_1+m_2}_{l}(\lambda)  \notag\\
    &\times |n_1+n_2-u;m_1+m_2-l\rangle_c|u;l\rangle_r
\end{align}

With the inverse transformation, the Coulomb matrix elements in Eq.~\eqref{eq:Coulomb_matrix_elements} can be written as:
\begin{align}
    &V_{1,2;3,4} = \sum_{\substack{u,l=0 \\ u',l'=0}}^{\substack{n_1+n_2,m_1+m_2 \\ n_3+n_4,m_3+m_4}} R^{n_1+n_2}_{u} R^{m_1+m_2}_{l} 
  R^{n_3+n_4}_{u'} R^{m_3+m_4}_{l'}(\lambda) \notag \\
  & \times \delta_{n_1+n_2-u,n_3+n_4-u'} \delta_{m_1+m_2-l,m_3+m_4-l'}\,\ {}_r\!\langle u;l|V_C|u';l'\rangle_r 
\end{align}

where ${}_r\!\langle u;l|V_C|u';l'\rangle_r $ is the Coulomb matrix elements in the relative basis. Note that the relative state in the center-of-charge basis differs from the one in quantum hall by scaling magnetic length $l_0$ to $l_0/\sqrt{1-\lambda^2}$( as shown in Eq.~\eqref{eq:ladder_operators_def}), thus the Coulomb matrix elements in the relative state is just scaled by $\sqrt{1-\lambda^2}$ from the one in quantum hall\cite{PhysRevB.87.245425}, and is given by:
\begin{align}\label{eq:Coulomb_relative}
    {}_r\!\langle u;l|V_C|u';l'\rangle_r &=  \sqrt{1-\lambda^2}\frac{\Gamma(|j|+\frac{1}{2})\Gamma(k'+\frac{1}{2})}{2|j|!}   \notag \\
     \times \sqrt{\frac{(k+|j|)!}{\pi (k'+|j|)!k'!k!}} &\, {}_3F_2\left(
\begin{array}{c}
 -k,\ |j|+\tfrac{1}{2},\ \tfrac{1}{2} \\
 |j|+1,\ \tfrac{1}{2}-k'
\end{array}
; 1 \right)
\end{align}
where $l,u,l',u' \geq0$  and $j=l-u$, $k=u+(j-|j|)/2$, $k'=u'+(j-|j|)/2$ and ${}_3F_2$ is generalized hypergeometric function.

For Coulomb interaction among opposite spin species $\sigma'=\bar \sigma$, due to the Bogoliubov type relation between the center-of-charge and individual state in Eq~\eqref{eq:bogoliiubov_opeartor_type}, the basis transformation does not take simple form as the same spin case. Therefore, we will use the same convention as in Ref~\cite{xu2025localizedexcitonslandaulevelmixing} and perform a particle-hole transformation on the spin $\bar\sigma$ species:
\begin{align}\label{eq:relative_basis_matrix_elements}
    V_{1,\bar2;3,\bar4}^{\sigma \bar\sigma} &= \langle n_1m_1\sigma,\bar n_2\bar m_2 \bar\sigma|V_C^{\sigma\bar\sigma}| n_3m_3\sigma,\bar n_4\bar m_4 \bar\sigma\rangle\\
    & = \langle n_1m_1\sigma, n_4 m_4\bar \sigma|V_C^{\sigma\bar\sigma}| n_3m_3\sigma, n_2 m_2\bar\sigma\rangle
\end{align}
So that the particle-hole state in the same bra and ket notation experiences the same direction of magnetic field, and one can use the same basis transformation in Eq.~\eqref{eq:basis_trans_appendix} to transfer the particle-hole state to center-of-charge and relative state, the Coulomb interaction only acts on the relative state the matrix elements in relative state is given by the same expression as in Eq.~\eqref{eq:Coulomb_relative}.

\section{RPA matrix}\label{sec:RPA_matrix}
In this appendix, we calculate the RPA matrix elements $A_{nn'}$ defined in Eq.~\eqref{eq:RPA_matrix_elements}.
We first transform the Landau-level indices from individual basis to center-of-charge and relative basis using Eq.~\eqref{eq:individual_to_COC} as the following:

\begin{align}
|0_p\uparrow,0_h\downarrow\rangle &= |0\rangle_c |0\rangle_r \\
|1_p\uparrow,0_h\downarrow\rangle 
     &= \sqrt{\frac{1+\lambda}{2}}|1\rangle_c |0\rangle_r +\sqrt{\frac{1-\lambda}{2}}|0\rangle_c|1\rangle_r\\|2_p\uparrow,0_h\downarrow\rangle &= \frac{1+\lambda}{2}|2\rangle_c |0\rangle_r+\sqrt{\frac{1-\lambda^2}{2}}|1\rangle_c|1\rangle_r +\frac{1-\lambda}{2}\notag \\
     &+|0\rangle_c|2\rangle_r
\end{align}
 Since the center-of-charge indices are conserved as indicated in Eq.~\eqref{eq:block_diagonal_V_COC}, the matrix elements in the scattering terms could be simplified into the matrix elements in the relative basis that is defined in Eq.~\eqref{eq:Coulomb_relative}.  Below are the explicit forms. For the diagonal elements, their form read as:
\begin{align}
    A_{00} &= \lambda + \kappa \sqrt{1-\lambda}\sqrt{\frac{\pi}{2}}- 0.326 g \lambda - \kappa\, {}_r\langle 0;j_0|V|0;j_0\rangle_r  , \\
    A_{11} &= 2\lambda+1+ \kappa \sqrt{1-\lambda}\sqrt{\frac{\pi}{2}}- 0.326g\lambda \notag \\
    &- \kappa\!\big[\frac{1+\lambda}{2}\,{}_r\langle 0;j_1|V|0;j_1\rangle_r
    \notag+ \frac{1-\lambda}{2}\,{}_r\langle 1;j_1|V|1;j_1\rangle_r\big]\\
    A_{22} &= 3\lambda+2+ \kappa \sqrt{1-\lambda}\sqrt{\frac{\pi}{2}}- 0.326g\lambda \notag \\
    &- \kappa \!\left[(\tfrac{1+\lambda}{2})^2{}_r\langle 0;j_2|V|0;j_2\rangle_r 
    + \tfrac{1-\lambda^2}{2}{}_r\langle 1;j_2|V|1;j_2\rangle_r  \right. \notag\\
    & \left.+ (\tfrac{1-\lambda}{2})^2{}_r\langle 2;j_2|V|2;j_2\rangle_r \right]
\end{align}

The off-diagonal elements only contain the scattering terms, and are given by the following:
\begin{align}
    A_{12} &= -\kappa \sqrt{\tfrac{1-\lambda}{2}}\,{}_r\langle 0;j_0|V|1;j_1\rangle_r, \quad A_{21} = A_{12}^*, \\
    A_{13} &= -\kappa \tfrac{1-\lambda}{2}\,{}_r\langle 0;j_0|V|2;j_2\rangle_r, \quad A_{31} = A_{13}^*, \\
    A_{23} &= -\kappa\!\left[ \tfrac{1+\lambda}{2}\sqrt{1-\lambda}\,{}_r\langle 0;j_1|V|1;j_2\rangle_r
      \right. \notag\\
    & \left.  + (\tfrac{1-\lambda}{2})^{3/2}\,{}_r\langle 1;j_1|V|2;j_2\rangle_r \right], \quad 
    A_{32} = A_{23}^*.
\end{align}

Here ${}_r\langle n;m|V|n';m'\rangle_r$ are the Coulomb matrix elements in the relative basis defined in Eq.~\eqref{eq:Coulomb_relative}. Because the Coulomb interaction conserves angular momentum, these matrix elements are nonzero only when $n-m = n' - m'$, which implies the spin-exciton modes at different Landau levels couple to each other only when $j_0 = j_1 - 1=j_2-2$.

\end{document}